\documentclass[prl,aps,twocolumn,superscriptaddress,showpacs,floatfix,amssymb,amsmath]{revtex4-2}

\usepackage{amsmath,amssymb,bm}
\usepackage{mathptmx}
\usepackage[T1]{fontenc}
\usepackage{microtype}
\usepackage{graphicx}
\usepackage[dvipsnames]{xcolor}
\usepackage{booktabs}
\usepackage[colorlinks=true,allcolors=MidnightBlue]{hyperref}

\hypersetup{
  pdftitle={Fold Catastrophe of a Hopf Texture Driven by a Navier–Stokes Blow-Up Analog},
  pdfauthor={Antti J. Niemi}
}

\newcommand{\bn}{\bm n}
\newcommand{\bU}{\bm U}

\newcommand{\bW}{\bm W}
\newcommand{\ba}{\bm a}
\newcommand{\bb}{\bm b}
\newcommand{\bOmega}{\bm\Omega}
\newcommand{\bv}{\bm v}

\newcommand{\bmn}{\mathbf m}
\newcommand{\bp}{\mathbf p}
\newcommand{\dx}{\Delta x}
\newcommand{\Cf}{C_{\rm f}}
\newcommand{\Qlat}{\mathcal Q_{\rm lat}}

\begin{document}

\title{
\LARGE%
Fold Catastrophes in a Hopf Texture Driven by \\ 
a Navier–Stokes Blow-Up Analog
}

\author{Antti J. Niemi}
\email{Antti.Niemi@su.se}
\affiliation{Nordita, Stockholm University,
Roslagstullsbacken 23, SE-106 91 Stockholm, Sweden}
\affiliation{Wilczek Quantum Center,  Shanghai Institute for Advanced Studies, 
			University of Science and Technology of China, Shanghai 201315, China}

\raggedbottom

\begin{abstract}
A two-component Bose--Einstein condensate can host a pseudospin Hopf texture coupled to the viscous normal 
flow of its thermal cloud. Motivated by OpenAI's reported finite-time Navier--Stokes singular solution, we evolve 
such a texture in a smooth  flow reproducing the geometry and similarity scaling of the collapsing 
axisymmetric core. The evolution creates and annihilates preimage pairs while preserving the Hopf charge. 
The first one-to-three transition exhibits the square-root opening
characteristic of a Thom $A_2$ fold. A subsequent three-to-five transition precedes a 
burst-like regime of recurrent fold catastrophes and near-caustic activity.
\end{abstract}
\maketitle

\textit{Introduction ---} OpenAI has recently reported a finite-time singular solution $\mathbf U_{\rm NS}(\mathbf x,t)$ of the
three-dimensional incompressible Navier--Stokes equation~\cite{OpenAI2026}. For each positive viscosity, the solution
starts from rest under a smooth force compactly supported in space and time. It remains smooth for $t<1$ and has uniformly
bounded kinetic energy, while its supremum norm diverges as $t\uparrow1$. We take these as reported properties without
assessing the construction, and ask how a physical texture would respond if a real flow developed a comparable collapsing core.

The leading singular region of the reported Navier--Stokes solution is akin to  an axisymmetric, double-celled vortical core. 
Fluid spirals inward and is expelled in opposite axial directions, producing two oppositely circulating meridional 
cells with a common swirl and a slight axial asymmetry. The swirl vanishes on the axis and is concentrated on a ring 
that contracts towards the origin as the blow-up time is approached. The radial scale shrinks faster than the axial 
one, while the peak velocity diverges. At the same time, the kinetic energy contained in the shrinking core tends to 
zero.

This co-swirling, counter-poloidal geometry is reminiscent of the geometry of a
Hopfion--antihopfion pair~\cite{Moffatt1969,Kamchatnov1982}, without
literally constituting one. Within each instantaneous profile, the 
helically winding streamlines lie on nested toroidal surfaces, evoking the fibre structure of a Hopfion, while 
reflection across the midplane preserves the swirl but reverses the poloidal circulation and hence the sign of the 
leading helicity density. In the reflection-symmetric limit, the two cells therefore have the parity of a neutral 
helicity dipole. The analogy is geometric rather than a statement of quantized
Hopf charge: the classical helicity of the velocity field is not generally quantized.

Because the reported velocity field is not available in a form suitable for direct simulation, we represent the leading 
core geometry by a divergence-free surrogate that retains the axisymmetric double-cell geometry, common swirl, 
contracting-ring structure, and anisotropic similarity scaling of the reported  core, together with a small axial 
asymmetry. At every finite time the prescribed field remains smooth, while its  characteristic scales contract 
and its peak velocity grows according to the singular similarity scaling as the blow-up time is approached.

Against this background, we place a genuine pseudospin Hopfion in the prescribed flow and follow its advection and
deformation as the core contracts. Its Hopf charge is conserved as long as the order parameter remains smooth, so the
texture cannot shed its topology as it is drawn towards the collapsing core, although its preimages may reorganize
locally. The helicity-dipole parity of the flow can thereby become imprinted on the preimage geometry, while the
increasing concentration drives a succession of charge-neutral fold events. This sequence can provide a topologically
resolved record of the flow's shrinking scales and increasing winding, helping to clarify the structure and physical
consequences of the proposed singularity without establishing its mathematical existence.

A binary Bose gas supplies the required order-parameter geometry: its relative amplitudes and phases define a unit
pseudospin texture, which carries an integer Hopf charge provided the condensate density remains nonzero and the
pseudospin approaches a fixed direction at the cloud boundary. Closely related three-dimensional skyrmions have been
proposed and studied in binary condensates~\cite{RuostekoskiAnglin2001,BattyeCooperSutcliffe2002}; projected onto the
pseudospin sphere, they yield Hopfions of the kind considered here. A Hopf-knot texture has also been created and
detected in the polar phase of a spin-1 $^{87}\mathrm{Rb}$ condensate~\cite{Hall2016}.

At finite temperature, the condensate coexists with a thermal cloud whose long-wavelength motion
becomes hydrodynamic in the collision-dominated regime~\cite{ZNG1999,HilkerEtAl2022}. The Bose fluid is compressible 
and supports first and second sound~\cite{HilkerEtAl2022}. However, when the flow
is slow compared with the sound speed and the density is uniform on the scale of the flow, 
density variations are of second order in the Mach number. The viscous mass-current mode of the cloud
is then described, to leading order, by the incompressible Navier--Stokes equation. 
We model this mode by our surrogate of the flow reported by OpenAI~\cite{OpenAI2026} and treat it as a
prescribed normal-flow protocol. A state-independent trap and drive primarily excite this common in-phase motion,
while spin drag suppresses relative motion between the two internal
components~\cite{NikuniWilliams2003,DuineStoof2009,Fava2018}; spin diffusion and thermal counterflow are therefore
neglected at leading order. In the absence of singular common-phase vortices, spatial twisting of the pseudospin
generates smooth condensate vorticity. This motivates coupling the texture to the rotational part of the normal flow
through an effective energy that penalizes a mismatch between the two vorticities. Although such a coupling does not
follow from standard neutral Gross--Pitaevskii theory, it is motivated by the extended Faddeev description of charged
two-condensate systems~\cite{BabaevFaddeevNiemi2002,BabaevFractionalFlux2002}, by related fractional-flux and
neutral-vortex structures in multicomponent superconductors~\cite{BabaevSpinSuperfluidity2005}, and by proposed
vorticity interactions in neutral multicomponent condensates~\cite{ChoKhimZhang2005}.

\emph{The blowup surrogate.---}
The explicit form  of the reported blowup flow $\mathbf U_{\rm NS}(\mathbf x,t)$ is not known, and our  
incompressible  surrogate $\mathbf U(\mathbf x,t)$, $\boldsymbol \nabla \cdot \mathbf U = 0$
is not obtained by solving the 
Navier--Stokes equations. It is constructed to reproduce only the leading axisymmetric 
double-cell geometry and  similarity scaling of the reported collapsing core.
With $\tau=1-t$ the time remaining to the singular time, following~\cite{OpenAI2026} 
we introduce the radial and axial similarity scales
\begin{equation*}
    \ell_r=\tau^{1/2},
    \qquad
    \ell_z=\tau^{1/2-h},
    \qquad
    R=\frac{r}{\ell_r},
    \qquad
    Z=\frac{z}{\ell_z},
\end{equation*}
with $h=0.005$.  Note that the radial scale contracts slightly faster than the axial one.
This concentration, combined with the velocity growth $\sim\tau^{-1/2-h}$, is the mechanism by which the 
flow can become pointwise singular at finite energy: the peak speed diverges while the
shrinking core occupies vanishing volume. The meridional flow is generated by the Stokes streamfunction
\begin{equation*}
    \psi
    =\tau^{-1/2-h}\ell_r^2
      R^2e^{-R^2/2}G(Z),
    \qquad
    G(Z)=(Z+j_0)e^{-Z^2/2},
\end{equation*}
where $j_0=0.05$ introduces the small axial asymmetry of the reported core. 
The corresponding velocity components are
\begin{equation*}
\begin{aligned}
    U_r
    &= -\frac{1}{r}\partial_z\psi
     = -\tau^{-1/2}R e^{-R^2/2}G'(Z), \\
    U_z
    &= \frac{1}{r}\partial_r\psi
     = \tau^{-1/2-h}(2-R^2)e^{-R^2/2}G(Z).
\end{aligned}
\end{equation*}
They produce radial inflow toward a dividing layer, opposite axial outflows, and radial return 
flow farther from the axis, giving two meridional circulation cells with opposite poloidal senses.
The common azimuthal swirl is prescribed as
\begin{equation*}
    U_\varphi
    =\tau^{-1/2-h}
      \frac{R}{(1+R^2)^{1+h}}
      \left(
        1+\frac{e^{-R^2/2}}{1+Z^2}
      \right).
\end{equation*}
It vanishes on the symmetry axis and is maximal on a ring of radius $r=O(\ell_r)$ that contracts toward the origin 
as $\tau\to0$. Thus the characteristic axial and azimuthal velocities grow as $\tau^{-1/2-h}$, while the radial component 
grows as $\tau^{-1/2}$. At the same time, the volume of the collapsing core shrinks sufficiently rapidly that its kinetic energy scales as
$ E_{\rm core}\sim  U^2\ell_r^2\ell_z \sim \tau^{1/2-3h}\longrightarrow0 $; this estimate applies to a bounded similarity region of axial
extent $O(\ell_z)$, not to the total energy or exterior completion of the prescribed finite-domain surrogate.
The surrogate therefore reproduces the defining concentration mechanism of the reported core:
the velocity diverges within a shrinking region whose contribution to the kinetic energy vanishes.
For the finite numerical domain, the streamfunction and swirl are suppressed near the outer boundary.
The nonaxisymmetric annular pulses and the detailed exterior completion of the reported construction are not retained.
Thus, $\mathbf U$ is a kinematic realization of the geometry and similarity scaling of the leading collapsing core,
rather than an independent Navier--Stokes blowup solution.

%
%
%
%
%
%
%
%
%

\textit{Effective pseudospin dynamics.---}
To implement our protocol in an effective model of a
binary Bose gas, we initialize its  pseudospin field
\(\bn\) as a smooth, relaxed Faddeev--Skyrme Hopfion of unit Hopf
charge \cite{Faddeev-Niemi,Battye-Sutcliffe,HietarintaSalo1999}.
We then evolve it under
the prescribed normal-flow drive, following its advection and
deformation as the collapsing core contracts. 
The underlying two-component condensate is parametrized as
$\bm\Psi=\sqrt{\rho_c}\,e^{i\Theta}Z,$ with $Z^\dagger Z=1$ and
$\bn=Z^\dagger\bm\sigma Z$ where \(\rho_c=\bm\Psi^\dagger\bm\Psi\) is 
the constant condensate
number density, \(\Theta\) is the common phase, and
\(Z=(z_\uparrow,z_\downarrow)^T\) is the normalized internal-state
spinor \cite{BabaevFaddeevNiemi2002} . This decomposition has the local U(1) redundancy
\(Z\rightarrow e^{i\chi}Z\), \(\Theta\rightarrow\Theta-\chi\) with
the associated \(CP^1\) connection and curvature
\begin{equation*}
a_i=-iZ^\dagger\partial_iZ,\qquad
b_i\equiv(\nabla\times\ba)_i
=\frac14\epsilon_{ijk}\,
\bn\cdot(\partial_j\bn\times\partial_k\bn).
\end{equation*}
For \(\bn\rightarrow\bn_\infty\) at spatial infinity, the Hopf charge is
\begin{equation*}
Q_{\rm H}=\frac{1}{4\pi^2}
\int\ba\cdot\bb\,d^3x\in\mathbb Z .
\end{equation*}
and in the London limit of the $SU(2)$-symmetric Gross--Pitaevskii theory we introduce
\begin{equation*}
s=\hbar \rho_c,\qquad
K=\frac{\hbar^2\rho_c}{m},\qquad
\kappa\equiv\frac{K}{s}=\frac{\hbar}{m},
\end{equation*}
where \(s\) is the Berry-term coefficient and \(K\) is the spin stiffness. The gauge-invariant condensate velocity is
$ \bv_s=\kappa(\nabla\Theta+\ba)$ and 
in the absence of common-phase vortices, the smooth pseudospin
texture carries condensate vorticity according to
\(\nabla\times\bv_s=\kappa\bb\). Within the fixed-density,
SU(2)-symmetric reduction, the quadratic spin-stiffness energy alone
does not stabilize a finite-size three-dimensional Hopf texture.
We therefore retain the quartic Faddeev--Skyrme curvature energy and
model its nondissipative coupling to the rotational part of the normal
flow by penalizing the mismatch between the two vorticities. Defining
the rescaled normal vorticity by $ \bOmega \equiv \kappa^{-1} \nabla\times\bU $
the minimal local positive locking form is
\begin{equation}
F_{\bU}[\bn]=\int d^3x\left[
\frac{K}{8}|\nabla\bn|^2+
\frac{M}{2}|\bb-\bOmega|^2\right],
\qquad M>0 .
\label{eq:energy}
\end{equation}
The  second term introduces the Faddeev--Skyrme stabilizing term, 
and favors energetically  local alignment of the condensate
and normal-flow vorticities.

To leading order, and with the normal flow imposed without backreaction, the effective action is
\begin{equation}
S_{\bU}=\int dt\,\left\{
\int d^3x\, s Z^\dagger
\bigl[i \partial_t+\bU\cdot(i\nabla)\bigr]Z
-F_{\bU}[\bn]\right\}.
\label{eq:action}
\end{equation}
The first term in Eq.~\eqref{eq:action} is the Berry-phase term written in the material coordinates of
the incompressible normal flow, so that the pseudospin precesses in the frame of the thermal cloud rather
than of the condensate. This constitutive assumption is analogous to the adiabatic spin-transfer torque
in ferromagnets~\cite{Bazaliy1998,ZhangLi2004}.
Variation at fixed \(\bU\), followed by projection onto the gauge-invariant pseudospin \(\bn\), gives
\begin{equation}
\bigl(\partial_t+\bU\cdot\nabla\bigr)\bn
=\frac{K}{2s}\bn\times\nabla^2\bn
-\frac{M}{s}
\left[\nabla\times(\bb-\bOmega)\right]_i\partial_i\bn .
\end{equation}
Using \(\nabla\cdot\bU=0\), and hence
\(\nabla\times\bOmega=-\nabla^2\bU/\kappa\), the part driven by the imposed vorticity can be absorbed 
into the effective transport velocity
\begin{equation}
\bW\equiv\bU-\frac{M}{s}\nabla\times\bOmega
=\bU+\ell_*^2\nabla^2\bU,\qquad
\ell_*^2\equiv\frac{M}{s\kappa}=\frac{M}{K}.
\label{eq:transport}
\end{equation}
The Laplacian term is a Fax\'en-like correction to local advection, with
$\ell_*$ setting the length scale over which spatial variations of the flow
affect the texture. 
The resulting closed evolution equation is the principal
theoretical result of our effective construction, and is the equation that we
integrate numerically:
\begin{equation}
\bigl(\partial_t+\bW\cdot\nabla\bigr)\bn
=\frac{K}{2s}\bn\times\nabla^2\bn
-\frac{M}{s}(\nabla\times\bb)_i\partial_i\bn .
\label{eq:dynamics}
\end{equation}

For a flow structure of width $L$, the Laplacian correction is of relative
order $(\ell_*/L)^2$. It becomes comparable to direct advection when
$L\sim\ell_*$ and can locally oppose, or for suitable profiles reverse, the
transport generated by $\bU$. This differential advection enables the
collapsing core to twist and fold the texture without simply carrying the
Hopfion bodily towards the core.
%
%
%
%
%
%
%
%
%
\begin{figure}[t]
  \centering
  \includegraphics[width=0.85\columnwidth]
  {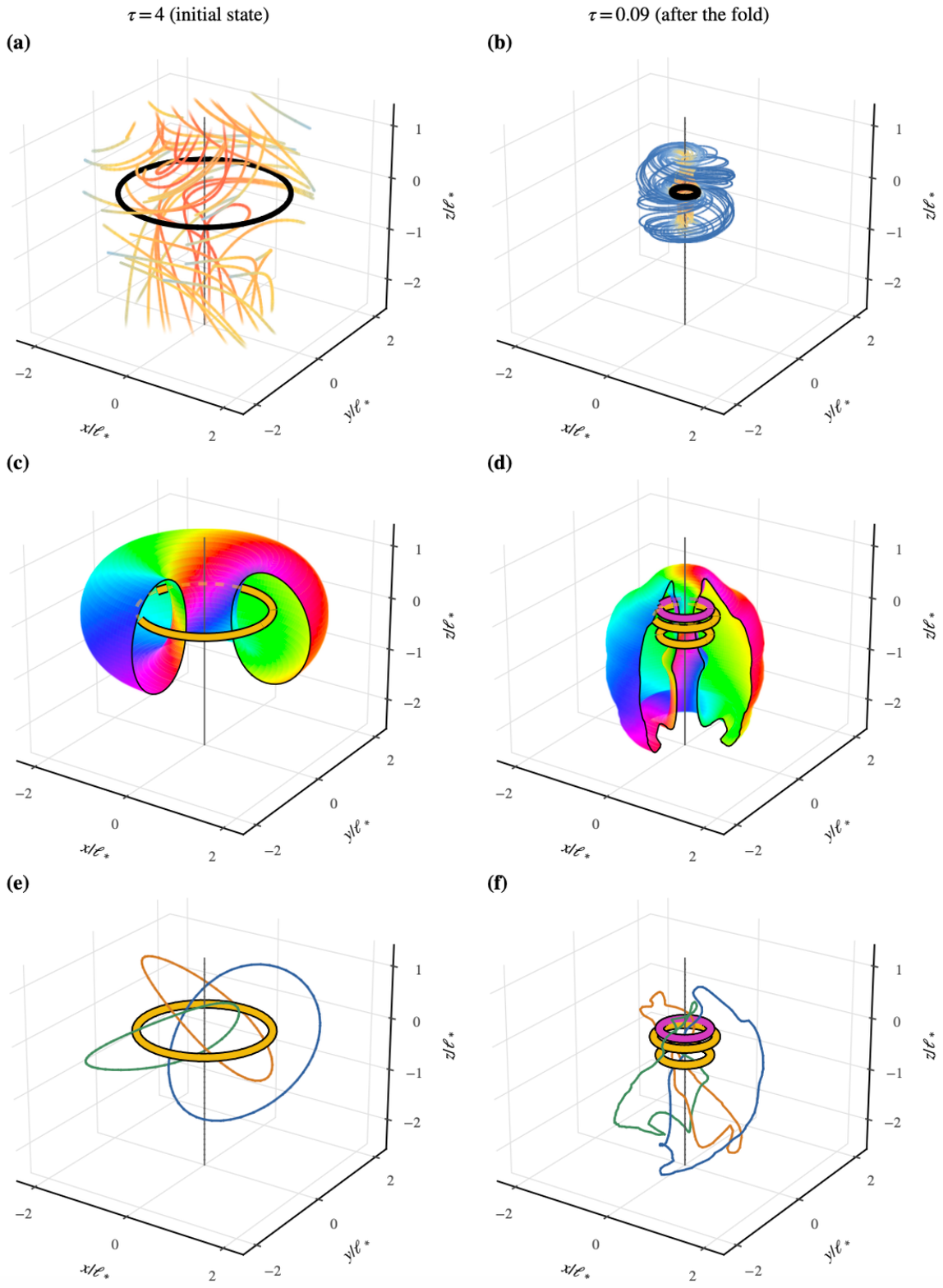}
  \caption{
All panels use the same spatial window, the same viewpoint and the same scale on all axes
(lengths in $\ell_*$). \textbf{(a,b)} The drive: streamlines of the surrogate core coloured
by angular rotation rate (blue slow, orange fast, normalised at each time); the black ring
marks the maximum of the swirl. At $\tau=4$ the core exceeds the window and only its inner
column is shown; at $\tau=0.09$ the whole core fits inside it.
\textbf{(c,d)} The Hopf texture: the equatorial surface $n_3=0$, cut open and coloured by the phase
$\arg(n_1+in_2)$, with the rings $\mathbf n=-\hat{\mathbf z}$ inside it; the relaxed Hopfion at 
$\tau=4$, and at $\tau=0.09$ the same texture drawn
toward the axis, stretched and twisted but not contracted. \textbf{(e,f)} Preimages of the
south pole: every ring $\mathbf n=-\hat{\mathbf z}$ coloured by its local degree (gold $+1$,
magenta $-1$), with the linked fibres of three equatorial points as thin curves. A single
ring at $\tau=4$; three coaxial rings at $\tau=0.09$ -- the original ring and a newborn pair
of opposite orientation on the fluid's dividing layer -- while $Q_H=1$ throughout. Between
the two times the drive has contracted sevenfold in radius; the texture has not followed it.}
\end{figure}
%
%
%
%
%
%
%
%
%
%
Equations~\eqref{eq:energy}--\eqref{eq:dynamics} define a phenomenological, nondissipative entrainment model rather 
than a generic consequence of standard finite-temperature Bose-gas theory. The material advection, quartic curvature 
term, and local $|\bb-\bOmega|^2$ locking are therefore model ingredients; mutual friction is not included.

\emph{Scale competition, localization, and twist.---}
Since the intrinsic terms in Eq.~\eqref{eq:dynamics} become 
enhanced on short length scales,  the singular growth of the effective 
transport does not need to imply that it overwhelms the texture dynamics.
For variations on the collapsing-core scale \(L\sim\ell_r\), and up to
profile-dependent factors, the deformation rate generated by
\(\mathbf W\) and the intrinsic texture-response rate scale as
\((U_\varphi/\ell_r)[1+\ell_\ast^2/\ell_r^2]\) and
\((\kappa/\ell_r^2)[1+\ell_\ast^2/\ell_r^2]\), respectively.
Their common short-scale enhancement therefore cancels in the ratio,
giving the core-scale estimate
\begin{equation*}
\Pi\equiv
\frac{\omega_{\rm drive}}{\omega_{\rm tex}}
=O\!\left(\frac{U_\varphi\ell_r}{\kappa}\right)
=O\!\left(\tau^{-h}\right).
\end{equation*}
Thus, although the drive becomes singular, its strength relative to
the intrinsic texture response grows only through the small exponent
\(h\).

We initialize the driven evolution at \(\tau_0=4\) with a relaxed
\(Q_{\rm H}=1\) Faddeev--Skyrme Hopfion \cite{Faddeev-Niemi,Battye-Sutcliffe,HietarintaSalo1999} 
($K=M=s=1$, with length measured in units of $\ell_\star$;  numerical
details are given in ~\cite{supplementary}.)
 As \(\ell_r/\ell_\ast\) decreases, the Laplacian term in
Eq.~\eqref{eq:transport} dominates the core-scale part of
\(\mathbf W\) and, for the prescribed profile, makes its inner radial
and azimuthal components oppose those of \(\mathbf U\). The meridional
projection of \(\mathbf W\) develops an elliptic fixed point, which
corresponds in the full three-dimensional flow to a circular
streamline. Numerically, the main Hopfion body remains localized on
the intrinsic scale \(r=O(\ell_\ast)\), rather than contracting with
the flow core.
This localization limits core-scale compression but does not remove
the differential rotation near the dividing layer. The latter
continues to wind the preimages and produces a layer at
\(r=O(\ell_r)\) in which folds develop. In the similarity coordinate
\(X=r/\ell_r\), this layer remains at \(X=O(1)\), whereas the main
Hopfion body lies at
\(X=O(\ell_\ast/\ell_r)\propto\tau^{-1/2}\). The lattice Hopf
diagnostic remains in the \(Q_{\rm H}=1\) sector within numerical
accuracy: The fold events described below are local
reorganizations of the preimage geometry rather than changes of the
global Hopf charge.
%
%
%
%
%
%
%
%
%
%

\emph{Fold criterion.---}
To identify a fold, we track the spatial preimage
$
\left\{\mathbf x:\bn(\mathbf x,\tau)=\bn_\ast\right\}
$
of a fixed target pseudospin direction as \(\tau\) varies. We write
\(\mathbf x=(\rho,\varphi,z)\) for a point in physical space and
\(\boldsymbol{\xi}=(\rho,z)\) for its meridional coordinates. The
texture is equivariant in the sense that a spatial rotation through \(\varphi\) about
the symmetry axis is accompanied by the same rotation about the third
pseudospin axis and the  full field is determined by its meridional profile
\(\mathbf m(\boldsymbol{\xi};\tau)\):
\begin{equation*}
\bn(\rho,\varphi,z;\tau)
 =R_z(\varphi)\mathbf m(\boldsymbol{\xi};\tau) \ \ \& \ \ 
f(\boldsymbol{\xi};\tau)
 =\left(
 \frac{m_1}{1-m_3},
 \frac{m_2}{1-m_3}
 \right)
\end{equation*}
where
\(\mathbf m(\boldsymbol{\xi};\tau)=(m_1,m_2,m_3)\), with the dependence
of each component on \((\boldsymbol{\xi};\tau)\) understood, and
\(R_z(\varphi)\) acts in pseudospin space. For fixed \(\tau\), we write
\(f_\tau(\boldsymbol{\xi})\equiv f(\boldsymbol{\xi};\tau)\).
 The map \(f_\tau\) gives stereographic coordinates
chosen so that the south-pole target
\(\bn_\ast=-\hat{\mathbf e}_3\) is represented by the origin. Thus
\(f_\tau(\boldsymbol{\xi})=0\) precisely when the meridional texture
takes the target value. Because \(\bn_\ast\) is invariant under
\(R_z(\varphi)\), every off-axis root lifts to an azimuthal preimage
ring in three dimensions; the nonaxisymmetric stability is not tested.

The critical set of the two-dimensional map \(f_\tau\) is
$
\mathcal C_\tau
 =\left\{
 \boldsymbol{\xi}:
 \det D_{\boldsymbol{\xi}}f_\tau(\boldsymbol{\xi})=0
 \right\},
$
and its image \(f_\tau(\mathcal C_\tau)\) is the caustic in the target
chart. A fold occurs when this caustic passes through the origin. At
the corresponding point \((\boldsymbol{\xi}_c,\tau_c)\) we have 
$f_{\tau_c}(\boldsymbol{\xi}_c)=0$ and  $J=D_{\boldsymbol{\xi}}f_{\tau_c}(\boldsymbol{\xi}_c),$
with   $\operatorname{rank}J=1.$
Under the generic nondegeneracy conditions below, this is a Whitney
fold, or Thom \(A_2\) catastrophe
\cite{Whitney1955,Arnold1992}.

Let \(\mathbf v\) and \(\mathbf u\) be unit right and left null vectors
of \(J\), $
J\mathbf v=0$ and $ \mathbf u^{\mathsf T}J=0$
and let \(q\) denote displacement along \(\mathbf v\). With
\(\mu=\tau_c-\tau\), elimination of the regular spatial direction
gives
\begin{equation*}
0=a q^2+b\mu
 +\mathcal O(q^3,\mu q,\mu^2),
\end{equation*}
where, with all derivatives evaluated at
\((\boldsymbol{\xi}_c,\tau_c)\),
\begin{equation*}
a =\frac{1}{2}\mathbf u^{\mathsf T}
 D_{\boldsymbol{\xi}}^2f_{\tau_c}
 [\mathbf v,\mathbf v],\ \ \ \& \ \ \
b=-\mathbf u^{\mathsf T}\partial_\tau f_\tau .
\end{equation*}
For a generic transverse crossing, \(a\neq0\) and \(b\neq0\). If
\(\boldsymbol{\xi}_\pm\) are the two nearby roots and $
d=\left|\boldsymbol{\xi}_+-\boldsymbol{\xi}_-\right| $
is their Euclidean separation in the meridional plane, then
\begin{equation*}
d^2=C_{\rm f}\mu+o(|\mu|),
\qquad
C_{\rm f}=-\frac{4b}{a}.
\end{equation*}
The branches exist on the side \(C_{\rm f}\mu>0\), giving the
universal square-root opening \(d\propto|\mu|^{1/2}\). Since
\(\mu=\tau_c-\tau\), \(C_{\rm f}>0\) describes pair creation as
\(\tau\) decreases, whereas \(C_{\rm f}<0\) describes annihilation.
The two roots have opposite local indices
\(\operatorname{sgn}\det D_{\boldsymbol{\xi}}f_\tau\) and therefore
form an index-neutral pair of preimage rings, not separate Hopfions
of opposite charge.

%
%
%
%
%
%
%
%

\vspace{0.2cm}

\begin{figure*}[t]
  \centering
  \includegraphics[width=0.8\textwidth]{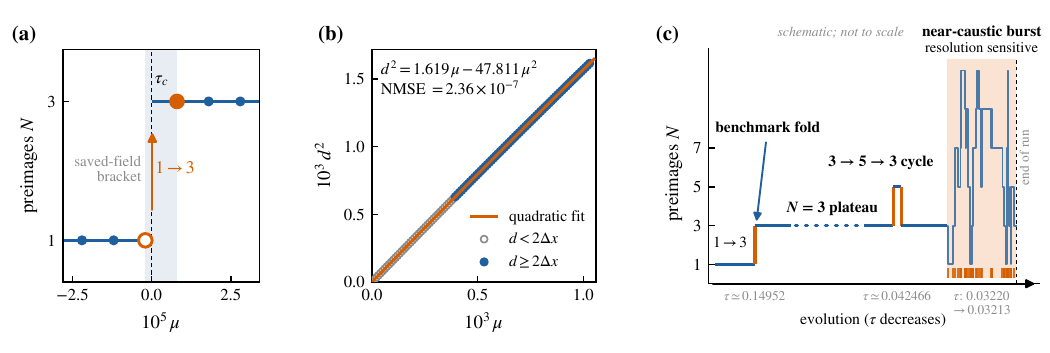}
  \caption{\textbf{From an isolated fold to late fold activity.}
  (a) Direct south-pole preimage count across the first pair-creation
  event at $\Delta x=0.0125$. The dashed line marks
  $\tau_c=0.1495185$, and the shaded interval is the
  saved-field bracket.
  (b) Squared separation of the newborn opposite-index pair versus
  $\mu=\tau_c-\tau$. 
  (c) Schematic summary, not to scale, of the separated activity windows: the
benchmark $1\to3$ fold, the $N=3$ plateau, a later converged $3\to5\to3$ fold
cycle, and a near-caustic window in which fold crossings burst and individual
events fall below the resolution of the computation.}
  \label{fig:first_fold_revised}
\end{figure*}

\emph{Identifying the first fold catastrophe.---}
After locating a transition on the $\Delta x=0.025$ survey lattice,
we restart shortly before it and repeat the evolution at the refined
spacing $\Delta x=0.0125$.  A subgrid analysis of the interpolated
fine-lattice trajectory resolves the first fold catastrophe. Two
adjacent saved fields directly bracket the change in the number $N$ of
south-pole preimages, with $N(0.14952)=1$ and $N(0.14951)=3.$
The signed preimage sum remains \(-1\) across the transition, so the
two new roots have opposite local indices and form a degree-neutral
pair. 
Solving the fold conditions within this interval gives
\(\tau_c\simeq0.1495185\), well inside the saved-field bracket.
Following the newborn roots along the post-fold branch we fit their
squared meridional separation to
\begin{equation}
d^2=1.619\,\mu-47.811\,\mu^2.
\label{d}
\end{equation}
The fit has normalized mean-square error (NMSE)
\(2.36\times10^{-7}\) over 102 saved post-fold fields spanning
$
8.53\times10^{-6}
\leq\mu\leq
1.019\times10^{-3}.
$
The positive coefficient of \(\mu\) gives the square-root opening
\(d\propto\sqrt{\mu}\) expected at a fold.
The directly resolved \(1\to3\) count change, the opposite local
indices of the newborn roots, and their stable square-root opening
together provide the characteristic numerical signature of a Thom
\(A_2\) pair-creation fold; see Fig.~\ref{fig:first_fold_revised}.

%
%
%
%
%
%
%
%
%
%

\vspace{0.2cm}

\emph{Late fold burst.---}
The benchmark $1\to3$ fold is followed by an extended sampled
$N=3$ plateau. Near $\tau\simeq0.04$, the continuations exhibit
short-lived additional-pair excursions. A selected $3\to5\to3$
cycle is locally identified on the interpolated trajectory as
Thom $A_2$ creation and annihilation of the same opposite-index
pair, with square-root opening and closing and agreement under
timestep halving.


The later sampled windows exhibit recurrent near-caustic activity; see
Fig.~\ref{fig:first_fold_revised}(c).  The $\Delta x=0.00625$
continuation reaches $\tau=0.03213$, whereas finer calculations
cover restricted onset and local-event intervals. The caustic
repeatedly approaches the south-pole target, while individual
crossings and transient pair separations become
resolution-sensitive. We therefore assign neither a unique
event-by-event ordering nor the benchmark square-root law to
every crossing. Taken together, the results support burst-like
neutral-pair creation and annihilation of preimages rather than their
monotonic accumulation. In all simulations the lattice diagnostic remains
$\mathcal Q_{\rm lat}=+1$, with no detected departure from the
unit Hopf sector, while the signed preimage sum is $-1$ in
degree-consistent snapshots.
Detailed numerical tests are given in the Supplemental Material.

\emph{Summary and outlook.---}
We find that a \(Q_{\rm H}=1\) Hopf texture, driven by a smooth
surrogate retaining the geometry and similarity scaling of the
collapsing core reported by OpenAI, undergoes local fold catastrophes
without changing its global topological sector.
The first \(1\to3\) transition that we observe has the characteristic numerical
signature of a Thom \(A_2\) fold: an opposite-index pair is created
with the expected square-root opening. Closer to the singular time, we observe
a burst of creation and annihilation events,  with several
preimage pairs transiently coexisting but with no discernible regularity
in the intervals between resolved events. Throughout these
reorganizations, the signed preimage sum remains \(-1\), and the
lattice Hopf diagnostic stays in the \(Q_{\rm H}=1\) sector.

The numerical evidence presented here concerns catastrophes of the
driven texture, not a singularity of the normal flow, and therefore
does not validate the reported Navier--Stokes construction. The
texture can nevertheless act as a topologically constrained recorder
of concentrated vortical motion. If a physical normal flow approached
a regime with comparable collapsing-core geometry and scaling, the
onset, localization, and space--time organization of its induced fold
events could provide experimentally accessible markers of that
approach.  Spin-resolved tomography could test whether concentrated
vortical forcing induces analogous preimage rearrangements,
without requiring the singular time to be reached.

\vspace{0.2cm}

\emph{Acknowledgements --}
 This research is
supported by the Swedish Research Council under Contract No. 2022-04037 
and by COST Action CA23134
(POLYTOPO).  
The author acknowledges extensive use of ChatGPT (OpenAI) and Claude (Anthropic)  
during the
development of this work. Dialogues with ChatGPT and Claude were used  in particular to clarify
aspects of the reported OpenAI Navier--Stokes construction, and to explore
the formulation of the prescribed normal-flow surrogate, 
and to assist with the preparation and revision of
Python code and manuscript text. All  simulations were performed locally with a Macbook Pro,
12 core M2 processors. All physical assumptions, mathematical
arguments, numerical procedures, and conclusions were initiated, independently
checked and accepted by the author, who takes full responsibility for
the results presented here.

\paragraph*{Data availability.---}
The numerical data, simulation and analysis code, parameter files,
and representative saved fields supporting the results of this work
will be deposited in a public repository. The complete high-resolution
simulation trajectories are available from the author upon reasonable
request.

\setlength{\bibsep}{1.2pt}
\begingroup
\raggedright

\endgroup

\newpage

\section{Appendix}
\section{Model parameters and initial state}
\label{sec:parameters}

The simulations use the fixed-total-density (London) effective pseudospin
model of the article. Length and time are measured in units
$\ell_*=\sqrt{M/K}$ and $t_*=\ell_*^2/\kappa$, where $\kappa=K/s$;
velocities and energies are measured in $\kappa/\ell_*$ and $K\ell_*$.
In dimensionless variables we set
\begin{equation}
 \begin{aligned}
 K=M=s&=1, & \kappa=\ell_*=t_*&=1,\\
 \nu&=1. &&
 \end{aligned}
 \label{eq:units}
\end{equation}
Here $\nu$ is the kinematic viscosity of the prescribed normal-flow
profile, measured in units of $\kappa$; thus the physical choice is
$\nu=\kappa$. The normal flow is prescribed, not obtained by integrating
Navier--Stokes dynamics. Evolution starts at $\tau_0=4$, with
$\tau=1-t$. A positive duration is therefore
$\Delta t=\tau_{\rm earlier}-\tau_{\rm later}$ in units of $t_*$.

The core profile uses $\ell_r=\tau^{1/2}$,
$\ell_z=\tau^{1/2-h}$, similarity exponent $h=0.005$, and axial bias
$j_0=0.05$. The overall swirl amplitude and its localized enhancement
are both set to unity ($e_0=\beta_{\rm sw}=1$). The similarity exponent
$h$ is distinct from the grid spacing $\dx$. On the numerical domain
$0\leq\rho\leq8$, $-8\leq z\leq8$, both the streamfunction and swirl
are multiplied by
\begin{equation}
 \begin{aligned}
 \chi(\rho,z)=\tfrac14&[1-\tanh(\rho-6)]\\
                       &\times[1-\tanh(|z|-6)] .
 \end{aligned}
 \label{eq:taper}
\end{equation}
The meridional velocity is obtained by differentiating the tapered
streamfunction, preserving incompressibility. The transport velocity is
$\bW=\bU+\ell_*^2\nabla^2\bU$, with the cylindrical vector Laplacian.
The untapered core is smooth for $\tau>0$; the implemented cutoff in
Eq.~\eqref{eq:taper} has a small midplane derivative mismatch of order
$\operatorname{sech}^2 6$. Its one-sided derivatives are evaluated on the
cell-centred axial grid.

The initial configuration is a \emph{numerically relaxed, unit-charge
Faddeev--Hopf soliton} of the no-flow Faddeev--Skyrme model, rather than
the unrelaxed Hopf ansatz. Starting from the standard unit-charge Hopf
spinor, the quadratic-plus-quartic energy is minimized without normal
flow on a larger domain, $R_{\rm box}=Z_{\rm box}=14$, using successively
finer relaxation meshes. The final relaxed profile has
$E_2\simeq34.119$, $E_4\simeq34.322$, and ring radius approximately
$1.229\ell_*$. This is a numerical soliton approximation, not an exact
continuum stationary solution. The profile is interpolated to the
production grid, normalized, and brought to the vacuum direction by a
smooth collar between coordinates $6.5$ and $7.5$, before applying the
drive. Subsequent refinements interpolate and normalize evolved fields
without re-relaxation.

\section{Numerical evolution and fold diagnostics}
\label{sec:diagnostics}

We impose the equivariance
$\bn(\rho,\varphi,z;t)=R_z(\varphi)\bmn(\rho,z;t)$ and integrate the
meridional reduction of the article's evolution equation. Azimuthal
advection contributes $-(W_\varphi/\rho)\hat{\mathbf e}_3\times\bmn$.
The lattice coordinates are $\rho_i=i\dx$ and
$z_j=-8+(j+\tfrac12)\dx$; the grids at
$\dx=0.0125,0.00625,0.003125$ contain, respectively,
$641\times1280$, $1281\times2560$, and $2561\times5120$ meridional sites.
Radial parity enforces axis regularity and outer ghost values are fixed
to the vacuum spin.

The intrinsic torque uses face gradients and geometric solid-angle
plaquette curvatures~\cite{BergLuscher}. Advection is second-order upwind;
time integration uses fourth-order Runge--Kutta with unit-spin projection
at every stage. Adaptive steps are limited by advection and spin stiffness.
The prescribed flow is interpolated quadratically within update blocks,
whose duration is at most $0.005\tau$ and which do not cross
saved-field times. The factor $\alpha$ in Table~\ref{tab:runs} multiplies
the stiffness timestep bound, not the saved-field spacing. No physical
damping is added, although upwinding and projection are not exactly
energy conserving. Local timestep comparisons start from identical
saved fields; spatial restarts retain their inherited history.

\begin{table*}[t]
\centering\small
\setlength{\tabcolsep}{5.5pt}
\begin{tabular}{@{}lcccc@{}}
\toprule
Calculation & $\dx$ & $\tau_{\rm start}\to\tau_{\rm end}$ & $\alpha$ & $\Delta\tau_{\rm save}$\\
\midrule
First-fold locator & $0.0125$ & $0.149520\to0.148500$ & $1/8$ & $10^{-5}$\\
Dense first-fold check & $0.0125$ & $0.149520\to0.149490$ & $1/8$ & $2\times10^{-7}$\\
Onset interval & $0.00625$ & $0.042500\to0.042450$ & $1,1/2$ & $10^{-6}$\\
Selected cycle & $0.00625$ & $0.042468\to0.042462$ & $1,1/2$ & $2\times10^{-7}$\\
Finer creation interval & $0.003125$ & $0.042475\to0.042468$ & $1$ & $10^{-6}$\\
Longer late window & $0.00625$ & $0.032200\to0.032130$ & $1$ & $10^{-6}$\\
Refined late onset & $0.003125$ & $0.032195\to0.0321888$ & $1$ & $10^{-7}$\\
\bottomrule
\end{tabular}
\caption{Analyzed intervals underlying the schematic stages in Fig.~2(c)
of the article. These local calculations do not constitute a single
uniformly refined trajectory. The first row lists the audited portion
of the locator output.}
\label{tab:runs}
\end{table*}

We monitor $E_2=\tfrac18\int|\nabla\bn|^2\,d^3x$,
$E_4=\tfrac12\int|\bb|^2\,d^3x$, neighbouring-spin angles, and the
meridional geometric diagnostic
$\Qlat=(4\pi)^{-1}\sum_{\square}\Omega_{\square}$, where
$\Omega_{\square}$ is the oriented solid angle of a plaquette.
Its reported value remains $+1$. This is a diagnostic of the equivariant
unit sector, not an independent evaluation of the signed
three-dimensional Hopf integral; comparison requires consistent
orientation conventions.

For the bicubic component interpolant $\bp$, the normalized south-pole
chart is $f=(p_1,p_2)/(|\bp|-p_3)$. Root searches solve
$F=(p_1,p_2)=0$ with $p_3<0$, which has the same target roots and local
indices. The signed census is
$\mathcal D=\sum_i\operatorname{sgn}\det Df(\boldsymbol\xi_i)$,
with baseline $\mathcal D=-1$. Whole-domain cycle checks combine
subdivision of candidate bicubic cells with Newton refinement and a
repeated search at greater subdivision depth. Unresolved cells remain
flagged. Agreement of the signed sum is a consistency check, not proof
that no neutral pair has been missed.

Between saved times, component fields are interpolated linearly in
$\tau$. A local fold satisfies $F=0$, $\operatorname{rank}DF=1$, and
nonzero coefficients
$a=\tfrac12u^{\mathsf T}D^2F[v,v]$ and
$b=-u^{\mathsf T}\partial_\tau F$, with unit left and right null vectors
$u,v$~\cite{Whitney1955,Arnold1992}. Its separation law is
\begin{equation}
 \begin{aligned}
 d^2&=\Cf(\tau_c-\tau)+o(|\tau_c-\tau|),\\
 \Cf&=-4b/a .
 \end{aligned}
 \label{eq:fold}
\end{equation}
Thus $\Cf>0$ describes creation and $\Cf<0$ annihilation as $\tau$
decreases. We additionally check opposite indices, local
absence/presence on the appropriate sides, and coalescence along the
null direction. Branch following connects observed roots to the
critical point and tests pair identity through a cycle. Local solver
precision is distinct from the accuracy of the evolved field and its
temporal interpolation.

\section{Benchmark and later fold examples}
\label{sec:folds}

\paragraph{Benchmark and sampled plateau.}
The counts $N(0.14952000)=1$ and $N(0.14951000)=3$ bracket the first
fold. The original local solve gives $\tau_c\simeq0.14951853$ and
$\Cf\simeq1.6339$. For 102 post-fold fields,
\begin{equation}
 \begin{aligned}
 d^2&=1.619003\mu-47.8109\mu^2,\\
 8.53\times10^{-6}&\leq\mu\leq1.019\times10^{-3},
 \qquad \mu=\tau_c-\tau .
 \end{aligned}
 \label{eq:benchmark}
\end{equation}
The recorded normalized mean-square error is $2.36\times10^{-7}$;
the leading coefficient changes by less than $0.08\%$ when the
nearest-grid pairs are excluded. The opposite-index pair and stable
opening identify the fold on this discrete trajectory. The benchmark
has not been repeated below $\dx=0.0125$.

A continuous $\dx=0.0125$ survey from $\tau=0.14850$ supports an
extended sampled $N=3$ plateau, with no detected return to one. Its
live census uses $0.01\leq\rho\leq1.2$, $-0.5\leq z\leq0.3$ and
samples at intervals $10^{-3}$ above $\tau=0.0425$ and $5\times10^{-5}$
below. Additional-pair activity appears near $\tau\simeq0.0422$.
These observations do not exclude shorter unsampled excursions or
establish a lifetime for an original $1\to3\to1$ cycle.

\paragraph{Selected creation--annihilation cycle.}
The $\dx=0.00625$ onset continuation contains several brief excursions.
In a selected $3\to5\to3$ cycle, the same opposite-index pair is born at
$\tau_c^{(+)}\simeq0.04246636$ and lost at
$\tau_c^{(-)}\simeq0.04246307$. Both endpoints pass the local $A_2$
tests, with $\Cf^{(+)}\simeq2.6110$ and
$\Cf^{(-)}\simeq-4.5361$, and the opening and closing fits support
Eq.~\eqref{eq:fold}. The lifetime is
$T=\tau_c^{(+)}-\tau_c^{(-)}\simeq3.28\times10^{-6}$; direct saved-field
brackets give $3.0\times10^{-6}\leq T\leq3.4\times10^{-6}$.
Halving the integration timestep gives agreement to the digits quoted
here. One interior global census remains flagged, but local following
connects the created pair to the annihilation. The maximum sampled
separation is $0.255\dx$, so this is a classification on the interpolated
trajectory. Moreover, reducing the saving interval from $10^{-6}$ to
$2\times10^{-7}$ shifts the reconstructed birth time by about
$1.85\times10^{-7}$; integration-step agreement alone does not bound
that reconstruction error.

\paragraph{Finer-grid creation.}
A separate continuation transfers the evolved $\dx=0.00625$ field at
$\tau=0.042475$ to $\dx=0.003125$. It has screened $N=3$ at
$0.042474$ and $N=5$ at the following six saved times through $0.042468$,
where the closest opposite-index separation is $1.67\dx$. On the
two-field interpolant across the creation bracket, the local tests give
$\tau_c\simeq0.04247371$, $\Cf\simeq10.23$, and
$(\rho_c,z_c)\simeq(0.335074,-0.141415)$.

Backward following of the observed roots and independently initialized
forward following connect both branches to this critical point;
halving the continuation increment leaves the connection unchanged
within the numerical tolerances. Near the fold, the narrowest reported
fit gives exponent $0.5001$ and $C_{\rm fit}/\Cf=1.00094$. This verifies
the branch association and opening on the same interpolant, not the
accuracy of linear interpolation between saved times. A corresponding
finer-grid annihilation and identification with the coarser cycle have
not been established; the same cycle and its parameters are therefore
not claimed to be spatially converged.

\section{Late near-caustic activity and scope}
\label{sec:late}

\begin{figure}[b]
\centering
\includegraphics[width=\columnwidth]{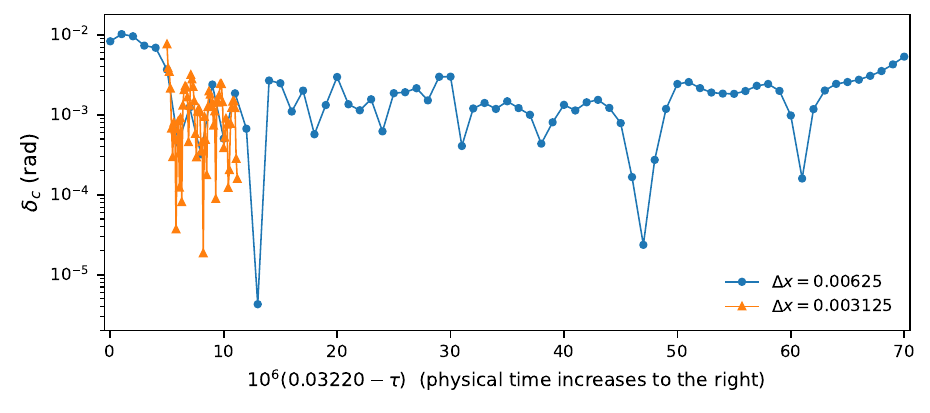}
\caption{Caustic-distance estimates for the late activity in Fig.~2(c) of
the article. Circles: 71 fields at $\dx=0.00625$ over
$0.03220000\geq\tau\geq0.03213000$, saved every $10^{-6}$.
Triangles: 63 fields at $\dx=0.003125$ over
$0.03219500\geq\tau\geq0.03218880$, saved every $10^{-7}$.
The vertical scale is logarithmic; physical time increases to the right.
Lines join saved values. Small $\delta_c$ indicates caustic proximity,
not by itself a resolved temporal crossing.}
\label{fig:caustic}
\end{figure}

Figure~\ref{fig:caustic} provides the measured-data counterpart to the
late window shown schematically in Fig.~2(c) of the article. We estimate
the minimum angular distance from the south-pole target to the caustic,
\begin{equation}
 \begin{aligned}
 \delta_c(\tau)&=\min_{\boldsymbol\xi\in C_\tau\cap\mathcal R}
 2\arctan|f_\tau(\boldsymbol\xi)|,\\
 C_\tau&=\{\det Df_\tau=0\},\\
 \mathcal R&=[0.01,1.2]\times[-0.5,0.3].
 \end{aligned}
 \label{eq:caustic}
\end{equation}
The normalized chart is used for both the critical set and the distance.
Contour-seeded searches at $\dx/4$ and $\dx/8$ retain multiple local
minima. They share seeds, so their agreement is not an independent
completeness or spatial-convergence test.

The caustic repeatedly approaches the target, typically within
$O(10^{-3})$ rad. At the seven common times, the coarser and finer
median distances are $1.24\times10^{-3}$ and $1.22\times10^{-3}$ rad,
respectively, although individual values differ appreciably. The finer
regional census is degree-consistent in 50 of 63 snapshots, with counts
between three and eleven; the remaining snapshots are flagged.
These data support recurrent near-caustic activity, not a converged
ordering of every crossing or a monotonic accumulation of pairs.

All results concern the equivariant, prescribed-flow model. Density
zeros, backreaction, mutual friction, and nonaxisymmetric stability are
not tested, and no completed boundary control excludes reflected waves.
Numerical loss of resolution is not evidence of a singular pseudospin
field. The accompanying data files retain the processed caustic series
and local-fold summaries; the article's Fig.~2(c) remains a schematic
rather than a complete, uniformly resolved trajectory.

\begin{thebibliography}{99}

\bibitem{OpenAI2026}
OpenAI, ``Finite Time Blowup for Navier--Stokes'' (2026),
\href{https://cdn.openai.com/pdf/32d9f210-8b73-45e0-91bc-82a30aef8a9a/navier-stokes.pdf}{OpenAI preprint}.

\bibitem{Moffatt1969}
H.~K. Moffatt, ``The degree of knottedness of tangled vortex lines,''
\emph{J. Fluid Mech.} \textbf{35}, 117--129 (1969),
\href{https://doi.org/10.1017/S0022112069000991}{doi:10.1017/S0022112069000991}.

\bibitem{Kamchatnov1982}
A.~M. Kamchatnov, ``Topological solitons in magnetohydrodynamics,''
\emph{Sov. Phys. JETP} \textbf{55}, 69--73 (1982).

\bibitem{RuostekoskiAnglin2001}
J.~Ruostekoski and J.~R. Anglin,
``Creating Vortex Rings and Three-Dimensional Skyrmions in Bose--Einstein Condensates,''
\emph{Phys. Rev. Lett.} \textbf{86}, 3934--3937 (2001),
\href{https://doi.org/10.1103/PhysRevLett.86.3934}{doi:10.1103/PhysRevLett.86.3934}.

\bibitem{BattyeCooperSutcliffe2002}
R.~A. Battye, N.~R. Cooper, and P.~M. Sutcliffe,
``Stable Skyrmions in Two-Component Bose--Einstein Condensates,''
\emph{Phys. Rev. Lett.} \textbf{88}, 080401 (2002),
\href{https://doi.org/10.1103/PhysRevLett.88.080401}{doi:10.1103/PhysRevLett.88.080401}.

\bibitem{Hall2016}
D.~S. Hall, M.~W. Ray, K.~Tiurev, E.~Ruokokoski, A.~H. Gheorghe, and M.~M\"ott\"onen,
``Tying quantum knots,''
\emph{Nat. Phys.} \textbf{12}, 478--483 (2016),
\href{https://doi.org/10.1038/nphys3624}{doi:10.1038/nphys3624}.

\bibitem{ZNG1999}
E.~Zaremba, T.~Nikuni, and A.~Griffin,
``Dynamics of Trapped Bose Gases at Finite Temperatures,''
\emph{J. Low Temp. Phys.} \textbf{116}, 277--345 (1999),
\href{https://doi.org/10.1023/A:1021846002995}{doi:10.1023/A:1021846002995}.

\bibitem{HilkerEtAl2022}
T.~A. Hilker, L.~H. Dogra, C.~Eigen, J.~A.~P. Glidden, R.~P. Smith, and Z.~Hadzibabic,
``First and Second Sound in a Compressible 3D Bose Fluid,''
\emph{Phys. Rev. Lett.} \textbf{128}, 223601 (2022),
\href{https://doi.org/10.1103/PhysRevLett.128.223601}{doi:10.1103/PhysRevLett.128.223601}.

\bibitem{NikuniWilliams2003}
T.~Nikuni and J.~E. Williams,
``Kinetic Theory of a Spin-1/2 Bose-Condensed Gas,''
\emph{J. Low Temp. Phys.} \textbf{133}, 323--375 (2003),
\href{https://doi.org/10.1023/A:1026206724886}{doi:10.1023/A:1026206724886}.

\bibitem{DuineStoof2009}
R.~A. Duine and H.~T.~C. Stoof,
``Spin Drag in Noncondensed Bose Gases,''
\emph{Phys. Rev. Lett.} \textbf{103}, 170401 (2009),
\href{https://doi.org/10.1103/PhysRevLett.103.170401}{doi:10.1103/PhysRevLett.103.170401}.

\bibitem{Fava2018}
E.~Fava, T.~Bienaim{\'e}, C.~Mordini, G.~Colzi, C.~Qu, S.~Stringari, G.~Lamporesi, and G.~Ferrari,
``Observation of Spin Superfluidity in a Bose Gas Mixture,''
\emph{Phys. Rev. Lett.} \textbf{120}, 170401 (2018),
\href{https://doi.org/10.1103/PhysRevLett.120.170401}{doi:10.1103/PhysRevLett.120.170401}.

\bibitem{BabaevFaddeevNiemi2002}
E.~Babaev, L.~D. Faddeev, and A.~J. Niemi,
``Hidden symmetry and knot solitons in a charged two-condensate Bose system,''
\emph{Phys. Rev. B} \textbf{65}, 100512(R) (2002),
\href{https://doi.org/10.1103/PhysRevB.65.100512}{doi:10.1103/PhysRevB.65.100512}.

\bibitem{BabaevFractionalFlux2002}
E.~Babaev,
``Vortices with Fractional Flux in Two-Gap Superconductors and in Extended Faddeev Model,''
\emph{Phys. Rev. Lett.} \textbf{89}, 067001 (2002),
\href{https://doi.org/10.1103/PhysRevLett.89.067001}{doi:10.1103/PhysRevLett.89.067001}.

\bibitem{BabaevSpinSuperfluidity2005}
E.~Babaev,
``Fractional-Flux Vortices and Spin Superfluidity in Triplet Superconductors,''
\emph{Phys. Rev. Lett.} \textbf{94}, 137001 (2005),
\href{https://doi.org/10.1103/PhysRevLett.94.137001}{doi:10.1103/PhysRevLett.94.137001}.

\bibitem{ChoKhimZhang2005}
Y.~M. Cho, H.~Khim, and P.~M. Zhang,
``Topological Objects in Two-Component Bose--Einstein Condensates,''
\emph{Phys. Rev. A} \textbf{72}, 063603 (2005),
\href{https://doi.org/10.1103/PhysRevA.72.063603}{doi:10.1103/PhysRevA.72.063603}.


\bibitem{Faddeev-Niemi}
L.~D. Faddeev and A.~J. Niemi,
``Stable knot-like structures in classical field theory,''
\emph{Nature} \textbf{387}, 58--61 (1997),
\href{https://doi.org/10.1038/387058a0}
{doi:10.1038/387058a0}.

\bibitem{Battye-Sutcliffe}
R.~A. Battye and P.~M. Sutcliffe,
``Knots as Stable Soliton Solutions in a Three-Dimensional Classical
Field Theory,''
\emph{Phys. Rev. Lett.} \textbf{81}, 4798--4801 (1998),
\href{https://doi.org/10.1103/PhysRevLett.81.4798}
{doi:10.1103/PhysRevLett.81.4798}.

\bibitem{HietarintaSalo1999}
J.~Hietarinta and P.~Salo,
``Faddeev--Hopf Knots: Dynamics of Linked Un-Knots,''
\emph{Phys. Lett. B} \textbf{451}, 60--67 (1999),
\href{https://doi.org/10.1016/S0370-2693(99)00054-4}
{doi:10.1016/S0370-2693(99)00054-4}.

\bibitem{Bazaliy1998}
Ya.~B. Bazaliy, B.~A. Jones, and S.-C. Zhang,
``Modification of the Landau--Lifshitz equation in the presence of a spin-polarized current in
colossal- and giant-magnetoresistive materials,''
\emph{Phys. Rev. B} \textbf{57}, R3213--R3216 (1998),
\href{https://doi.org/10.1103/PhysRevB.57.R3213}{doi:10.1103/PhysRevB.57.R3213}.

\bibitem{ZhangLi2004}
S.~Zhang and Z.~Li,
``Roles of Nonequilibrium Conduction Electrons on the Magnetization Dynamics of Ferromagnets,''
\emph{Phys. Rev. Lett.} \textbf{93}, 127204 (2004),
\href{https://doi.org/10.1103/PhysRevLett.93.127204}{doi:10.1103/PhysRevLett.93.127204}.

\bibitem{supplementary}
See Appendix  for details of the numerical initialization,
integration, and diagnostics.

\bibitem{Whitney1955}
H.~Whitney,
``On Singularities of Mappings of Euclidean Spaces. I. Mappings of the
Plane into the Plane,''
\emph{Ann. Math.} \textbf{62}, 374--410 (1955),
\href{https://doi.org/10.2307/1970070}
{doi:10.2307/1970070}.

\bibitem{Arnold1992}
V.~I. Arnold, \emph{Catastrophe Theory}, 3rd ed.
(Springer-Verlag, Berlin, 1992),
\href{https://doi.org/10.1007/978-3-642-58124-3}
{doi:10.1007/978-3-642-58124-3}.

\end{thebibliography}

\begin{thebibliography}{9}
\bibitem{BergLuscher}
B.~Berg and M.~L\"uscher, ``Definition and statistical distributions of a
topological number in the lattice $O(3)$ sigma-model,''
\emph{Nucl. Phys. B} \textbf{190}, 412--424 (1981),
\href{https://doi.org/10.1016/0550-3213(81)90568-X}{doi:10.1016/0550-3213(81)90568-X}.
\bibitem{Whitney1955}
H.~Whitney, ``On Singularities of Mappings of Euclidean Spaces. I. Mappings of
the Plane into the Plane,'' \emph{Ann. Math.} \textbf{62}, 374--410 (1955),
\href{https://doi.org/10.2307/1970070}{doi:10.2307/1970070}.
\bibitem{Arnold1992}
V.~I.~Arnold, \emph{Catastrophe Theory}, 3rd ed. (Springer-Verlag, Berlin, 1992),
\href{https://doi.org/10.1007/978-3-642-58124-3}{doi:10.1007/978-3-642-58124-3}.
\end{thebibliography}
\end{document}